\documentclass[twocolumn,article,showpacs,amsmath,amssymb]{revtex4}
\usepackage{graphicx}

\begin{document}

\title{The CSM extension for the description of positive and negative parity bands in even-odd nuclei}

\author{A. A. Raduta$^{a),b),c)}$ and C. M. Raduta$^{b)}$}

\address{$^{a)}$ Department of Theoretical Physics and Mathematics, Bucharest University, Bucharest, POBox MG11, Romania}

\address{$^{b)}$ Department of Theoretical Physics, Institute of Physics and Nuclear Engineering, Bucharest, POBox MG6, Romania}

\address{$^{c)}$ Academy of Romanian Scientists, 54 Splaiul Independentei, Bucharest 050094, Romania}

\begin{abstract}A particle-core Hamiltonian is used to describe the lowest parity partner bands $K^{\pi}=1/2^{\pm}$ in $^{237}$U and $^{239}$Pu.
The quadrupole and octupole boson  Hamiltonian  associated to the core is identical to the one  previously used for the description of four positive and four negative parity bands in the neighboring even-even isotopes. The single particle space for the odd nucleon consists of three spherical shell model states, two  of positive and one of negative parity. The particle-core Hamiltonian consists of four terms: a quadrupole-quadrupole, an octupole-octupole, a spin-spin and a rotational $\hat{I}^2$ interaction, with $\hat {I}$ denoting the total angular momentum. The parameters involved in the particle-core coupling Hamiltonian were fixed by fitting four particular energies in the two bands. The calculated  excitation energies are compared with the corresponding experimental data as well as with those obtained with other approaches. Also, we searched for some signatures for a static octupole deformation in the considered odd isotopes.   
\end{abstract}


\maketitle

The coherent state model (CSM)\cite{Rad81} describes in a realistic fashion three interacting bands, ground, beta and gamma, in terms of quadrupole bosons. The formalism was later extended \cite{Rad97,Rad02,Rad03,Rad003,Rad06,Rad006}
 by considering the octupole degrees of freedom. The most recent extension describes eight rotational bands, four of positive  and four of negative parity. Observables like excitation energies, intraband E2 and interband E1, E2 and E3 reduced transition probabilities have been calculated and the results were compared with the corresponding experimental data. The formalism works well for both
near spherical and deformed nuclei excited in low and high angular momentum states. Indeed, we considered all states with $J\le 30$ in 
both, the positive and the negative parity bands. Signatures for a static octupole deformation in ground as well as in excited bands have been pointed out in several even-even nuclei.

The aim of this paper is to extend CSM for the even-odd nuclei which exhibit both quadrupole and octupole deformation.
Thus, we consider a particle-core Hamiltonian:
\begin{equation}
H=H_{sp}+H_{core}+H_{pc},
\end{equation}
where  $H_{sp}$ is a spherical shell model Hamiltonian associated to the odd nucleon, while $H_{core}$ is a phenomenological Hamiltonian which describes the collective motion of the core in terms of quadrupole and octupole bosons. This term is identical to that used in Ref.\cite{Rad006} to describe eight rotational bands in even-even nuclei. The two subsystems interact with each other by $H_{pc}$, which has the following expression:
\begin{eqnarray}
H_{pc}=&-&X_2\sum_{\mu}r^2Y_{2,-\mu}(-)^{\mu}\left(b^{\dagger}_{2\mu}+(-)^{\mu}b_{2,-\mu}\right)\nonumber\\
       &-&X_3\sum_{\mu}r^3Y_{3,-\mu}(-)^{\mu}\left(b^{\dagger}_{3\mu}+(-)^{\mu}b_{3,-\mu}\right)\nonumber\\
       &+&X_{jJ}\vec{j}\vec{J}+X_{I^2} \vec{I}^2.
\end{eqnarray}
 $b^{\dagger}_{\lambda \mu}$ denotes the components of the  $\lambda$-pole (with $\lambda$=2,3) boson operator.
The term $\vec{j}\vec{J}$ is similar to the spin-orbit interaction from the shell model and expresses the interaction between the angular momenta of the odd-particle and the core. The last term is due to the rotational motion of the whole system, $\vec{I}$ denoting the total angular momentum of the particle-core system.

The core states are described by eight sets of mutually orthogonal functions, obtained by projecting out the angular momentum and the parity from four quadrupole and octupole deformed functions: one is a product of two coherent states:
\begin{equation}
\Psi _{g}=e^{ f(b_{30}^{+}-b_{30})}e^{d(b^+_{20}-b_{20})}|0\rangle _{2}|0 \rangle _{3} \equiv \Psi_{o}\Psi_{q}
|0\rangle _{2} | 0\rangle _{3},
\end{equation}
while the remaining three are polynomial boson excitations of $\Psi_{g}$. The parameters $d$ and $f$ are real numbers and simulate the quadrupole and octupole deformations, respectively. The vacuum state for the k-pole boson, $k=2,3$, is denoted by $|0\rangle_{k}$.

The particle-core interaction generates a deformation for the single particle trajectories. Indeed, averaging the model Hamiltonian with
 $\Psi_{g}$, one obtains a deformed single particle Hamiltonian, $H_{mf}$ which plays the role of the mean field for the particle motion:
\begin{equation}
H_{mf}={\cal C}+H_{sp}-2dX_2 r^2Y_{20}-2fX_3 r^3Y_{30},
\end{equation}
where ${\cal C}$ is a constant determined by the average of $H_{core}$. The Hamiltonian $H_{mf}$ represents an extension of the  Nilsson Hamiltonian by adding the octupole deformation term. In Ref.\cite{Rad99} we have shown that in order to get the right deformation dependence of the single particle energies $H_{mf}$ must be amended with a monopole-monopole interaction, $M\omega^2r^2\alpha_{00}Y_{00}$, where the monopole coordinate $\alpha_{00}$ is determined from the volume conservation restriction. This term has a constant contribution within a band. The constant value is, however, band dependent.

In order to find the eigenvalues of the model Hamiltonian we follow several steps:

1) In principle the single particle basis could be determined by diagonalizing $H_{mf}$ amended with the monopole interaction. The product basis for particle and core may be further used to find the eigenvalues of $H$. Due to some technical difficulties in restoring the
rotation and space reversal symmetries for the composite system wave function, this procedure is however tedious and therefore we prefer a simpler method.
Thus,  the single particle space consists of three spherical shell model states with angular momenta 
 $j_1,j_2, j_3$. We suppose that $j_1$ and $j_2$ have the parity $\pi=+$, while $j_3$ has a negative parity $\pi=-$. 
Due to the quadrupole-quadrupole interaction the odd particle from the state $j_1$ can be promoted to $j_2$ and vice-versa. The octupole-octupole interaction connects the states $j_1$ and $j_2$ with $j_3$. Due to the above mentioned effects the spherical and space reversal symmetries of the single particle motion are broken. Thus, instead of dealing with a spherical shell model state coupled to a deformed core without reflection symmetry, as the traditional particle-core approaches proceed, here the single particle orbits are lacking the spherical and space reversal symmetries and by this, their symmetry properties are consistent with those of the phenomenological core.        

2) We remark that  $\Psi_{g}$ is a sum of two states of different parities. This happens due to the specific structure of the octupole coherent state: 
\begin{equation}
\Psi_{o}=\Psi_{o}^{(+)}+\Psi_{o}^{(-)}.
\end{equation}
The states of a given angular momentum and positive parity  can be obtained through projection from the intrinsic states:
\begin{equation}
|n_1l_1j_1 K\rangle\Psi^{(+)}_{o}\Psi_{q},\;\;|n_2l_2j_2 K\rangle\Psi^{(+)}_{o}\Psi_{q},\;\;|n_3l_3j_3 K\rangle\Psi^{(-)}_{o}\Psi_{q}.
\end{equation}
The projected states of negative parity  are obtained from the states:
\begin{equation}
|n_1l_1j_1 K\rangle\Psi^{(-)}_{o}\Psi_{q},\;\;|n_2l_2j_2 K\rangle\Psi^{(-)}_{o}\Psi_{q},\;\;|n_3l_3j_3 K\rangle\Psi^{(+)}_{o}\Psi_{q}.
\end{equation}
The angular momentum and parity projected states are denoted by:
\begin{eqnarray}
\varphi^{(+)}_{JM}(j_iK;d,f)&=&N^{(+)}_{i;JK}P^{J}_{MK}|n_il_ij_iK\rangle \Psi^{(+)}_{o}\Psi_{q},i=1,2\nonumber\\
\varphi^{(+)}_{JM}(j_3K;d,f)&=&N^{(+)}_{3;JK}P^{J}_{MK}|n_3l_3j_3K\rangle \Psi^{(-)}_{o}\Psi_{q},\nonumber\\
\varphi^{(-)}_{JM}(j_iK;d,f)&=&N^{(+)}_{i;JK}P^{J}_{MK}|n_il_ij_iK\rangle \Psi^{(-)}_{o}\Psi_{q},i=1,2\nonumber\\
\varphi^{(-)}_{JM}(j_3K;d,f)&=&N^{(+)}_{3;JK}P^{J}_{MK}|n_3l_3j_3K\rangle \Psi^{(+)}_{o}\Psi_{q}.
\end{eqnarray}
For the quantum number $K$ we consider the lowest three values, i.e. $K=1/2,3/2,5/2$.
Note that the earlier particle-core approaches \cite{RaCea,Lea} restrict the single particle space to a single $j$, which results in
eliminating the contribution of the octupole-octupole interaction. 

3) Note that for a given  $j_i$, the projected states with different $K$ are not orthogonal.
Indeed, the overlap matrices :

\begin{eqnarray}
A^{(+)}_{K,K'}(j_l;d,f)&=&\langle \varphi^{(+)}_{JM}(j_lK;d,f)|\varphi^{(+)}_{JM}(j_lK';d,f)\rangle,\nonumber\\
                   l&=&1,2,3;\;K,K'=1/2,3/2,5/2,
\nonumber\\
A^{(-)}_{K,K'}(j_l;d,f)&=&\langle \varphi^{(-)}_{JM}(j_lK;d,f)|\varphi^{(-)}_{JM}(j_lK';d,f)\rangle, 
\nonumber\\
                  l&=&1,2,3;\;K,K'=1/2,3/2,5/2,
\end{eqnarray} 
are not diagonal.
By diagonalization, one obtains the eigenvalues $a^{(\pm)}_p(j_l)$ and the corresponding eigenvectors $V^{(\pm)}_K(j_l,p)$, with $K=1/2,3/2,5/2$ 
and $p=1,2,3$.
Then, the functions:

\begin{eqnarray}
\Psi^{(+)}_{JM}(j_l,p;d,f)=N^{(+)}_{l;Jp}\sum_{K}V^{(+)}_{K}(j_l,p)\varphi^{(+)}_{JM}(j_lK;d,f),
\nonumber\\
\Psi^{(-)}_{JM}(j_l,p;d,f)=N^{(-)}_{l;Jp}\sum_{K}V^{(-)}_{K}(j_l,p)\varphi^{(-)}_{JM}(j_lK;d,f).\nonumber\\
\label{psiplmi}
\end{eqnarray}
are mutually orthogonal. The norms are given by:
\begin{equation}
\left(N^{(\pm)}_{l;Jp}\right)^{-1}=\sqrt{a^{(\pm)}_p(j_l)}.
\end{equation}
For each of the new states, there is a term in the defining sum (10), which has a maximal weight. The corresponding  quantum number $K$ is conventionally  assigned to the mixed state. 

4) In order to simulate the core deformation effect on the single particle motion, in some cases the projected states corresponding to different $j$ must be mixed up.

\begin{eqnarray}
\Phi^{(+)}_{JM}(p;d,f)=\sum_{l=1,2,3}{\cal A}^{(+)}_{pl}\Psi^{(+)}_{JM}(j_lp;d,f),\nonumber\\
\Phi^{(-)}_{JM}(p;d,f)=\sum_{l=1,2,3}{\cal A}^{(-)}_{pl}\Psi^{(-)}_{JM}(j_lp;d,f).
\end{eqnarray}
The amplitudes ${\cal A}^{(\pm)}_{pl}$ can be obtained by diagonalizing
 $H_{mf}$.
	
The energies of the odd system are approximated by the average values of the model Hamiltonian corresponding to the projected states:
\begin{eqnarray}
E^{(+)}_J(p;d,f)=\langle \Phi^{(+)}_{JM}(p;d,f)|H|\Phi^{(+)}_{JM}(p;d,f)\rangle,\nonumber\\
E^{(-)}_J(p;d,f)=\langle \Phi^{(-)}_{JM}(p;d,f)|H|\Phi^{(-)}_{JM}(p;d,f)\rangle.
\label{energ}
\end{eqnarray}

The matrix elements involved in the above equations can be analytically calculated. They have been used to calculate the excitation energies for one positive and one negative parity bands in two odd isotopes: $^{237}$U si $^{239}$Pu. The parameters defining $H_{core}$, as well as the deformation parameters $d$ and $f$ are those used to describe the properties of  eight rotational bands in the even-even neighboring isotopes. The single particle states are spherical shell model states with the appropriate parameters for the $(N,Z)$ region of the considered isotopes \cite{Ring}. Our calculations correspond to the single particle states: $(j_1,j_2,j_3)=(2g_{7/2},2g_{9/2},1h_{9/2})$. 
In the expansion (12) a small admixture of the states $(j_1;j_3)$ and $(j_2;j_3)$ was considered: $|{\cal A}^{(+)}_{i,3}|^2=0.04,\; |{\cal A}^{(-)}_{i,3}|^2=0.06, i=1,2$. The mixing amplitude of the third state is negligible. Energies
(\ref{energ}) depend on the interaction strengths $X_2,X_3, X_{jJ}$ and $X_{I^2}$.
These were determined by fitting four particular energies in the two bands of different parities. The results of the fitting procedure are given in Table I. Inserting these in Eqs.\ref{energ} the energies in the two bands are readily obtained. The theoretical results for excitation energies, listed in Table II, agree quite well with the corresponding experimental data.
To have a measure for the agreement quality, we calculated the r.m.s. values for the deviations of the calculated values from the experimental ones. The results for $^{237}$U and $^{239}$Pu are 48.97 keV and 31.8 keV, respectively.

\begin{table}
\begin{tabular}{|c|cc|}
\hline
Parameters           &   $^{237}$U      &   $^{239}$Pu\\
\hline                                   
$X_2b^2$[keV]         &1.080            &-2.515 \\
$X_3b^3$[keV]         &2.227           &4.937\\
$X_{jJ}$[keV]      &-5.817           &-3.985 \\
$X_{I^2}$[keV]     &4.634            &5.050  \\
\hline
\end{tabular}
\caption{Parameters involved in the particle-core Hamiltonian obtained by fitting four excitation energies. Here $b$ denotes
the oscillator length: $b=(\frac{\hbar}{M\omega})^{1/2};\,\hbar\omega =41A^{-1/3}$. The usual notations for nucleon mass (M) and atomic number (A) were used}.
\end{table}
\begin{table}
\begin{tabular}{|c|cc|cc|cc|cc|}
\hline
    &\multicolumn{4}{c|}{$^{237}$U}&\multicolumn{4}{c|}{$^{239}$Pu}\\
\hline
    & \multicolumn{2}{c|}{$\pi =+$}&\multicolumn{2}{c|}{$\pi= -$}
& \multicolumn{2}{c|}{$\pi =+$}&\multicolumn{2}{c|}{$\pi= -$}\\
    \hline
J     &    Exp.    &   Th          &   Exp.     &      Th. &
Exp. & Th. & Exp. & Th. \\ 
\hline
1/2  &0.0         &  0.0          &            &398.5
&0.0            &0.0          &469.8         &469.8\\
3/2  &11.4        &11.4           &            &454.4
&7.9            &7.9          &492.1         &477.7\\
5/2  &56.3        &74.6           &            &475.5
&57.3           &62.8         &505.6         &498.3\\
7/2  &82.9        &106.9          &            &550.3
&75.7           &108.4        &556.0         &549.8\\
9/2  &162.3       &191.2          &            &581.3
&163.8          &183.5        &583.0         &572.0\\
11/2 &204.1       &231.8          &            &680.9
&193.5          &222.0        &661.2         &655.2\\
13/2 &317.3       &347.7          &            &721.9
&318.5          &338.1        &698.7         &685.7\\
15/2 &375.1       &393.1          &846.4       &846.4
&359.2          &386.5        &806.4         &799.9\\
17/2 &518.2       &544.2          &930.0       &899.1
&519.5          &534.9        &857.5         &839.5\\
19/2 &592.0       &592.0          &1027.5      &1046.6
&570.9          &592.2        &992.5         &984.2\\
21/2 &762.8       &780.3          &1131.0      &1113.3
&764.7          &773.7        &1058.1        &1033.3\\
23/2 &853.0       &829.0          &1250.7      &1281.3
&828.0          &839.2        &1219.4        &1208.3\\
25/2 &1048.7      &1065.8         &1376.1      &1364.8
&1053.1         &1054.4       &1300.9        &1267.2\\
27/2 &1155.1      &1108.8         &1515.7      &1550.2
&1127.8         &1127.8       &1487.4        &1472.2\\
29/2 &1372.2      &1378.3         &1662.3      & 1654.0
&1381.5         &1377.0       &1584.9        &1541.2\\
31/2 &1494.1      &1421.6         &1821.8      &1852.8
 &1467.8         &1458.0       &1795.4        &1776.0\\
33/2 &1729.2      &1728.7         &1987.7      &1981.0
&1748.5         &1744.2       &1908.9        &1855.4\\
35/2 &1868.2      &1772.5         &2166.5      &2188.9
&1847.0         &1831.3       &2143.4        &2119.8\\
37/2 &2117.2      &2117.2         &2349.7      &2346.1
 &2152.2         &2150.2       &2272.0        &2209.8\\
39/2 &2272.2      &2161.7         &2547.5      &2558.3
&2263.0         &2245.0       &2529.4        &2503.6\\
41/2 &2530.1      &2544.1         &2746.7      &2749.4
 &2590.1         &2597.9       &2672.0        &2604.4\\
43/2 &2702.5      &2589.4         &2960.5      &2960.5
&2714.0         &2700.5       &2951.4        &2927.5\\
45/2 &2963.8      &3009.5         &3174.7      &3191.3
&3060.1         &3087.5       &3108.0        &3039.3\\
47/2 &3154.5      &3055.6         &3401.5      &3395.3
 &3198.0         &3198.0       &3407.0        &3395.3\\
49/2 &3415.8      &3513.7         &3630.0      &3671.7
 &3559.1         &3619.1       &3578.0        &3514.4\\
51/2 &3625.5      &3560.5         &3865.0      &3862.4
 &3713.0         &3737.0       &3895.0        &3895.8\\
53/2 &3886.8      &4057.8         &4105.0      &4190.9
&4087.1         &4194.0       &4080.0        &4029.9\\
55/2 &4115.0      &4104.8         &4344.0      &4350.0
&4256.0         &4319.8       &4413.0        &4436.7\\
\hline
\end{tabular}
\caption{Excitation energies in
$^{237}$U  and $^{239}$Pu, for the bands characterized by $K^{\pi}=\frac{1}{2}^{+}$ and  $K^{\pi}=\frac{1}{2}^{-}$ respectively,   
 are given in keV. The results of our calculations (Th.) are compared with the corresponding experimental data (Exp.) Data are from Ref.\cite{Zhu}.}
\end{table}
\begin{figure}[ht!]
\begin{center}
\includegraphics[width=0.4\textwidth]{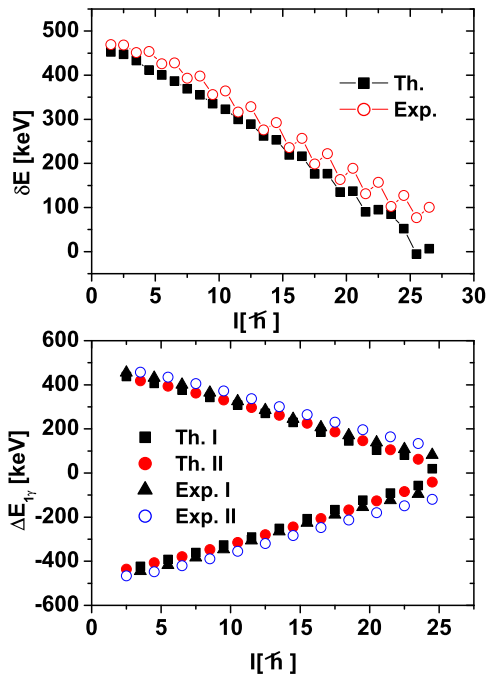}
\end{center}
\caption{The energy displacement functions characterizing  the isotope $^{239}Pu$, are plotted as a function of angular momentum.
Experimental data are taken from Ref.\cite{Zhu}. In the lower panel, the symbols labeled by $I$ correspond to the the  of states
which includes $1/2^+$ as the lowest $(I-2)$ state in Eq.(14). When the lowest state $(I-2)$ in (14) is the state $1/2^-$, the associated symbols to the results and data bear the label $II$.  }
\label{Fig. 1}
\end{figure}
Further we addressed the question whether one could identify signatures for static octupole deformation in the two bands.
To this goal we plotted the energy displacement functions \cite{Rad02,Rad03,Bona0}:
\begin{eqnarray}
\delta E(I)&=&E(I^-)-\frac{(I+1)E((I-1)^+)+IE((I+1)^+)}{2I+1},\nonumber\\
\Delta_{1,\gamma}(I)&=&\frac{1}{16}[6E_{1,\gamma}-4E_{1,i\gamma}(I_1)-4E_{1,\gamma}(I+1)\\
&&+E_{1,\gamma}(I-2)+E_{1,\gamma}(I+2)],\nonumber\\
E_{1,\gamma}&=&E(I+1)-E(I).
\end{eqnarray}
The first function, $\delta E$ vanishes when the   excitation energies of the parity partner bands depend linearly on
$I(I+1)$ and, moreover, the moments of inertia of the two bands are equal. Thus, the vanishing value of $\delta E$ is considered to be a signature for octupole deformation. If the excitation energies depend quadratically on $I(I+1)$ and the coefficients of the $[I(I+1)]^2$ terms for the positive and negative parity bands are equal, the second energy displacement function $\Delta E_{1,\gamma}$ vanishes, which again suggests that a static octupole deformation shows up. The parities associated to the angular momenta, involved in $\Delta E_{1,\gamma}$ 
are as follows: the levels $I$ and $I\pm 2$ have the same parity, while levels $I$ and $I\pm 1$ are of opposite parities.
The results plotted in Figs. 1,2 suggest that a static octupole deformation is possible for the states with angular momenta $I\ge \frac{25}{2}$ belonging to the two parity partner bands. 

The spectra of the odd isotopes, considered here, have been previously studied in Refs.\cite{Zub,Deni,Bona} using a quadrupole-octupole Hamiltonian in the intrinsic  deformation variables $\beta_2$ and $\beta_3$ separated in a  kinetic energy, a potential energy term and a Coriolis interaction. Due the specific structure of the model Hamiltonian, an analytical solution for the excitation energies in the two bands of opposite parities was possible. The agreement obtained in our approach for $^{239}$Pu is better than that shown in Ref.\cite{Bona}. However, the results from Ref.\cite{Bona} for $^{237}$U  agree better, with the corresponding data, than ours. Indeed, the r.m.s. values for the deviations of theoretical results, reported in Ref.\cite{Bona} from experimental data are 30 keV and 60 keV for $^{237}$U and $^{239}$Pu,  which are to be compared with 48.9 and 31.8 keV respectively, obtained with our approach.

Before closing, we would like to add few remarks about the possible development of the present formalism. Choosing for the core unprojected states, the generating states for the parity partner bands with $K^{\pi}=0_{\beta}^{\pm},2_{\gamma}^{\pm}, 1^{\pm}$ states, otherwise keeping the same single particle basis for the odd nucleon, the present formalism can be extended to another six bands, three of positive and three of negative parity. Another noteworthy remark refers to the chiral symmetry \cite{Frau} for the composite particle and core system. Indeed, in Ref.\cite{Rad006} we showed that starting from a certain total angular momentum of the core, the angular momenta carried by the quadrupole ($\vec{J}_2$) and octupole ($\vec{J}_3$)  bosons respectively, are perpendicular on each other. Naturally, we may ask ourselves whether there exists a strength for  the particle-core interaction such that the angular momentum of the odd particle becomes perpendicular to the plane ($\vec{J}_2, \vec{J}_3$). This would be a signature that the three component system exhibits a chiral symmetry. These issues are under work in our group and the results will be reported elsewhere.

As a final conclusion, one may say that the present CSM extension to odd nuclei can describe quite well the excitation energies in the
parity partner bands with $K^{\pi}=\frac{1}{2}^{\pm}$.

{\bf Acknowledgment.} This paper was supported by the Romanian Ministry of Education and Research under the contracts PNII, No. ID-33/2007
and ID-1038/2009.

\end{document}